**Dual-channel metasurfaces for refractometric and spatial heterogeneity sensing**

Zhongjun Jiang, Nicolas Gomez, Soyaib H Sohag and You Zhou*

*Department of Physics and Optical Science, University of North Carolina at Charlotte, Charlotte, NC, USA, 28223*

*Email: yzhou33@charlotte.edu

## Abstract

Nanophotonic sensors amplify analyte-induced optical signals through resonantly enhanced light-matter interactions. While effective for refractometric sensing, conventional sensing platforms typically collapse spatial variations in analyte distributions into ensemble-averaged readouts, thereby obscuring local heterogeneity. Here we present a dual-channel metasurface sensor capable of detecting both net analyte loading and nanoscale spatial inhomogeneity at deep-subwavelength scales. The device supports two distinct high-$Q$ modes with deeply structured near-field profiles of opposite parity. Their complementary modal responses decouple analyte perturbations into bulk and differential components, which are encoded into two spectrally resolvable channels. We demonstrate these sensing modalities across analyte configurations exhibiting diverse forms of nanoscale heterogeneity, achieving a refractometric sensitivity of $357\ \mathrm{nm\ RIU^{-1}}$ and inhomogeneity detection below λ/10. This dual-channel capability, which extends nanophotonic sensing from analyte quantity to its spatial variation, opens a new avenue for multidimensional analysis of biological and material systems.

Optical molecular sensors, which transduce the presence of surface-bound analytes into measurable optical signatures, provide important analytical tools for bioanalysis and medical diagnostics[1–3]. Nanophotonic sensors enable enhanced optical transduction by strengthening light–matter interactions through the excitation of optical resonances[3–5], offering a route toward non-destructive, label-free, and ultrasensitive molecular analysis. These strategies have been widely explored across plasmonic[6–13] and dielectric[14–21] platforms and are highly effective for transforming subtle refractive index changes into pronounced spectroscopic signatures. When analyte composition or morphology varies spatially, however, conventional readouts typically collapse these nanoscale variations into an ensemble-averaged response, leaving the underlying spatial heterogeneity unresolved. Accessing this hidden information would open a new analytical dimension for characterizing complex biological and material systems.

Nanophotonic sensing is fundamentally governed by the near-field modal profiles supported by photonic nanostructures. The spatial distributions and vector components of these modes determine the analyte properties to which a sensor responds. In prevailing sensing strategies, nanostructures are designed to support modes that concentrate electric field intensity into localized hotspots. When the modal profile varies sharply over the characteristic length scale of the analyte heterogeneity, different analyte regions are sampled with distinct optical weights, providing the spatial selectivity required for differential sensing. Although deep-subwavelength field localization has been achieved[9,12,22,23], the associated sensing modes typically encode analyte perturbations through spectral features dominated by net refractive loading. Consequently, analytes with similar mean loading but different spatial distributions can collapse into the same readout channel (Fig. 1a). Separating these contributions therefore motivates the engineering of two key modal features that provide (i) deeply structured near fields for differential sampling and (ii) an independent readout channel that selectively encodes spatial heterogeneity.

In this work, we develop a dual-channel metasurface sensor capable of detecting both refractometric loading and spatial inhomogeneity in analyte distributions at deep-subwavelength scales. As illustrated in Fig. 1b, the sensing response arises from two high-quality-factor ($Q$) modes with deeply structured near-field profiles that respond to complementary regimes of analyte perturbation. The first mode supports an odd-parity profile with respect to the central mirror plane that responds to net analyte coverage, providing a refractometric measure of the bulk response. The second mode originates from a bound state in the continuum (BIC) with an even-parity field distribution. Transverse variations render this mode radiative, yielding a spectrally resolved inhomogeneity signature. The two channels thereby decompose analyte perturbations into the mean ($\bar{n}$) and differential ($\nabla n$) components, enabling both to be independently retrieved from the spectroscopic readout.

To enable these nontrivial modal features, we employ our recently developed inverse design framework[24,25], which directly optimizes the near-field resonant hotspots in photonic nanostructures. Conventional nanophotonic design, which typically relies on empirically identified, simply shaped nanostructures guided by physical intuition, has yielded libraries of heuristic device templates[14,26–30]. However, these approaches offer limited control over complex near-field landscapes, including the realization of arbitrarily structured mode profiles and their extension to multifunctional, multi-resonant responses. In contrast, our framework directly formulates the desired near-field modal properties as figures of merit (FoMs) and optimizes them through efficient exploration of a vast freeform design space. Motivated by the need to probe nanoscale spatial variations, we focus on compressing the lateral field variation of the sensing mode, in contrast to conventional strategies that primarily target resonant $Q$ factors or local field enhancement. We first target the odd-parity resonance introduced in Fig. 1b, characterized by a deeply structured, vertically antiparallel mode profile. We monitor the complex $\boldsymbol{E_z}$ fields at two closely spaced probe positions, $\boldsymbol{r_1}$ and $\boldsymbol{r_2}$, inside the nanostructures and define the FoM as their field difference:

$$\text{FoM} = |\boldsymbol{E_z}(\boldsymbol{r_1}) - \boldsymbol{E_z}(\boldsymbol{r_2})| \tag{1}$$

Maximizing this FoM therefore yields the targeted odd-parity mode configuration with a large field contrast compressed into a deep-subwavelength region. Interestingly, although only this mode is explicitly formulated as the design target, we later find that the optimized structure supports an additional even-parity BIC mode structured on the same deep-subwavelength scale, therefore giving rise to the distinct inhomogeneity sensing channel. The emergence of this mode highlights the exploratory potential of inverse design: beyond realizing prescribed objectives, we uncover new optical responses that inherit shared modal characteristics within the same freeform nanostructure.

The metasurface sensor consists of a periodic array of meta-atoms arranged on a square lattice with a period of 920 nm and is designed to operate at a wavelength of 1.58 µm. We choose a 462-nm-thick silicon layer as the all-dielectric material to support multipolar Mie-type resonances. The design process is illustrated in Fig. 2a. Near the center of the unit cell, we place two near-field probes separated by 60 nm ($<\lambda/10$) to measure the local $\boldsymbol{E_z}$ fields. The FoM defined in Eq. (1) is then maximized under $x$-polarized plane wave incidence through topology optimization based on the adjoint-variable method[31–33], an iterative gradient-based approach for freeform nanophotonic design. In each iteration, far-field and near-field sources are used in the forward and adjoint simulations, respectively. The resulting electric fields are obtained to calculate the gradient of the FoM with respect to the permittivity distribution, which is then used to update the device topology until the optimization converges to a local optimum (Supplementary Section 1).

We employ the open-source finite-difference time-domain package MEEP[34], coupled with the Adam optimizer implemented in PyTorch[35], to perform gradient-based optimization. Mirror symmetry is imposed on the design space to enforce a transversely symmetric layout. To facilitate facile fabrication, minimum feature size and curvature constraints are applied during optimization to promote smooth curvilinear geometries[33,36–39]. Fig. 2b shows the evolution of the FoM during optimization, together with top-view permittivity profiles at selected intermediate iterations. The design is initialized with a grayscale material distribution and gradually evolves into a binary freeform geometry composed of silicon and air. The top-view $Re(E_z)$ distribution of the final structure (Fig. 2c, left) exhibits the targeted antiparallel configuration, characterized by a deep-subwavelength field variation and odd parity across the $yz$ mirror plane. To quantify the spatial extent of this field variation, we take a one-dimensional horizontal line cut through the structural center (white dashed line), revealing a sharp field reversal confined to a 0.08λ-wide region (Fig. 2c, right). We note that the optimized dielectric nanostructure geometrically resembles the widely studied bowtie antennas[22,40–43] that are well-known in the plasmonics community. Despite this top-view resemblance, the dielectric structure has a substantially larger vertical aspect ratio and supports multimode resonances dominated by vertical electric fields confined within the dielectric, in contrast to the in-plane dipolar field localized in the air gap of the plasmonic bowtie (see Supplementary Section 2 for detailed modal analysis).

Our metasurface sensor supports two distinct resonant channels with complementary responses to analyte perturbations. The first channel responds predominantly to net dielectric loading in the superstrate, providing a refractometric readout. To characterize this channel, we cover the metasurface with a uniform overlayer with $n = 1.7$ (Fig. 2d) and analyze the resulting resonant responses through multipolar decomposition, with the multipole moments calculated from the induced current density distributions within the nanostructures[44]. The decomposition (Fig. 2e) reveals a high-$Q$ hybrid resonance dominated by magnetic-dipole (MD) and electric-quadrupole (EQ) contributions. These multipolar contributions have been studied as a route to high-$Q$ resonances[45,46], with the radiative coupling governed by the relative strengths of the dipolar and quadrupolar components. The side-view $Re(E_z)$ distribution of this mode (Fig. 2f) reveals the odd-parity antiparallel configuration explicitly targeted during optimization. Its spectral response is dominated by net analyte loading and thus provides the refractometric sensing channel. To quantify this response, we vary the superstrate index from $n = 1$ to $n = 1.5$ and track the resonance wavelength. As shown in Fig. 2g, the resonance exhibits a linear redshift with increasing $n$, along with a variation in the $Q$ factor as the substrate–superstrate environment is modified (see Supplementary Section 3 for detailed $Q$-factor analysis). The refractometric sensitivity, defined as $S = \Delta\lambda_{\text{res}}/\Delta n$, reaches $357\ \text{nm/RIU}$, where RIU denotes refractive-index unit. The corresponding sensing $\text{FoM}_{\text{RI}}$[47], defined as $\text{FoM}_{\text{RI}} = S/\text{FWHM}$, where FWHM denotes the full-width at half-maximum of the resonance, is $115\ \text{RIU}^{-1}$ near the environmental index of 1.5.

The second channel is associated with a symmetry-protected BIC that responds selectively to perturbations with lateral variations. This mode remains dark in a laterally uniform environment but becomes radiative when analyte inhomogeneity breaks the transverse mirror symmetry. Because this BIC mode is intrinsically nonlocal, coherent lattice coupling requires the symmetry perturbation to be periodically repeated across the unit cells[48–52]. To validate this channel, we cover the metasurface with a 560-nm-thick dielectric layer exhibiting a transversely varying refractive-index profile within each unit cell. As shown in Fig. 2h, the index increases linearly from $n_{\text{left}}$ to $n_{\text{right}}$ along the $x$ direction with a variation of $\Delta n = n_{\text{right}} - n_{\text{left}}$ while maintaining a fixed mean index $\bar{n} = 1.7$. The multipolar decomposition in Fig. 2i reveals an additional mode dominated by the toroidal-dipole (TD) contribution, while the MD–EQ refractometric resonance is preserved owing to the same average dielectric loading. Interestingly, although this mode was not explicitly targeted during optimization, it emerges with a distinct even-parity field distribution characterized by two in-phase, vertically oriented field lobes structured on the same deep-subwavelength scale, as revealed by the $Re(E_z)$ distribution in Fig. 2j (see detailed field analysis of the TD mode in Supplementary Section 4). The resulting resonance provides a spectrally isolated signature of spatial inhomogeneity, with its $Q$ factor enabling quantitative assessment of the transverse index contrast. We further vary the linear index profile by changing the index contrast $\Delta n$ between its two ends, while keeping the mean index fixed, and extract the corresponding $Q$ factor. As shown in Fig. 2k, the $Q$ factor decreases monotonically with increasing $\Delta n$, following an inverse-square dependence, $Q \propto (\Delta n)^{-2}$, as expected for a symmetry-protected quasi-BIC.

Our sensor can detect spatial inhomogeneity in analyte distributions at deep-subwavelength dimensions owing to the deeply structured mode profile. To quantify this capability, we confine the analyte to the λ/10-wide gap between the two opposing tips of the nanostructure, within which the refractive index varies linearly along the $x$ direction (Fig. 3a, left). The transmittance spectra for laterally uniform ($\nabla_x n = 0$) and laterally varying ($\nabla_x n \neq 0$) analyte profiles with the same average index $\bar{n}$ are shown in Fig. 3a, right. The two configurations exhibit nearly identical refractometric signatures, whereas a pronounced inhomogeneity signature appears only for the laterally varying case despite the ultrasmall analyte volume. The magnitude of the transverse gradient controls the degree of mirror symmetry breaking and is encoded in the $Q$ factor of the inhomogeneity mode. We further decrease the end-to-end index contrast $\Delta n$ from 3.2 to 2.8 while keeping the same mean index and calculate the resonant near-field distributions at the two sensing wavelengths. As shown in Fig. 3b, the side-view $Re(E_z)$ distributions at the refractometric resonance remain nearly unchanged as the gradient magnitude decreases, whereas the inhomogeneity mode exhibits progressively enhanced near-field intensity.

We next extend the analysis to analyte configurations exhibiting diverse forms of nanoscale heterogeneity. The metasurface sensor is covered by a 500-nm-thick dielectric layer with a refractive index of 1.48, providing a planar surface above the nanostructures for analyte loading (Fig. 3c). We consider three representative analyte configurations: (1) a dielectric wedge with an inclined interface, representing continuous transverse variation; (2) a nanoscale grating, representing discrete geometric patterning; and (3) a pair of nanospheres with different refractive indices, representing compositional contrast. Corresponding laterally uniform or compositionally symmetric counterparts are analyzed as controls. Detailed geometric and material parameters for all analyte cases are provided in Supplementary Section 5. The corresponding transmittance spectra are shown in Fig. 3d–f. In each case, the heterogeneous analyte and its control exhibit nearly overlapping refractometric responses, whereas the inhomogeneity signature emerges only in the presence of transverse heterogeneity.

This unique dual-channel sensing modality originates from the two distinct high-$Q$ resonances that effectively decouple $\bar{n}$ and $\nabla n$ responses, which are otherwise conflated in a single averaged readout in conventional sensing schemes. To highlight this distinction, we evaluate the same analyte configurations using conventional refractometric sensing platforms. We first consider a plasmonic bowtie antenna formed by two opposing gold nanotriangles separated by a nanoscale gap[43]. The corresponding geometry and near-field distribution (Fig. 3g) reveal the characteristic field localization within the gap. The reflection spectra (Fig. 3h) show nearly overlapping signatures for the heterogeneous cases and their controls, indicating a response governed predominantly by the mean perturbation. We further consider a widely studied dielectric sensing architecture comprising a zigzag array of anisotropic resonators[15,27,53] (Fig. 3i, inset). The resulting spectra likewise show an ensemble-averaged single-channel response. Similar analyses using other high-$Q$ quasi-BIC metasurfaces are provided in Supplementary Section 6.

We experimentally validate the sensing concept by fabricating the designed metasurface in a 462-nm-thick amorphous-silicon film on a fused-silica substrate. The silicon film is deposited by plasma-enhanced chemical vapor deposition, and the freeform nanostructures are patterned using electron-beam lithography followed by reactive ion etching (see fabrication details in Methods). A scanning electron microscopy image of the fabricated device is shown in Fig. 4a, revealing well-defined nanoscale features consistent with the design. We first characterize its refractometric response by coating the device with a laterally uniform dielectric overlayer. Rather than varying the constituent index of the analyte, we spin-coat poly(methyl methacrylate) (PMMA; $n = 1.48$) films of different thickness onto the sensor (Fig. 4b). The subwavelength-thick PMMA forms a thin overlayer that progressively fills the interstitial volume surrounding the nanostructures, thereby providing a tunable analyte system with an effective index evolving from $n_{\text{eff}} = 1$ (uncoated) toward $n_{\text{eff}} = 1.48$ (complete filling).

Fig. 4c presents the simulated transmittance spectra for three representative PMMA thicknesses ranging from 180 to 260 nm, showing a progressive redshift of the resonance dips as the thickness increases. Experimentally, we vary the PMMA film thickness by adjusting the spin-coating speed from 2,000 to 4,000 rpm. Between successive measurements, the existing PMMA layer is thoroughly removed with acetone before a new film is applied. The resulting films remain thinner than the 462-nm-thick silicon nanostructures, thus reproducing the partial-filling regime considered in the simulations. The far-field transmittance spectra are measured under $x$-polarized illumination from a supercontinuum source, with the transmitted light collected by a custom-built imaging system and relayed to a high-resolution spectrometer (see measurement details in Methods). As shown in Fig. 4d, the measured transmittance spectra show well-resolved resonance dips under all spin-coating conditions.

Lastly, we experimentally demonstrate the inhomogeneity sensing modality using transversely varying dielectric perturbations. To coherently excite the nonlocal dark mode, we realize the periodic transverse perturbation as a PMMA wedge array over the sensor. The fabrication process is illustrated in Fig. 5a. A planar PMMA film is first deposited over the nanostructures, after which the substrate is tilted during reactive ion etching to sculpt the film into a slanted interface. We use oxygen plasma to selectively etch the PMMA while minimizing damage to the silicon nanostructures. The transverse variation of the resulting dielectric profile is controlled by adjusting the substrate tilt angle and etching depth. Fig. 5b shows the atomic force microscopy (AFM) image of the PMMA-covered sample after tilted etching. The corresponding three-dimensional surface rendering and one-dimensional line profile through the center (green dashed line) confirm the formation of a slanted PMMA interface. As a control, a laterally uniform overlayer is prepared using the same PMMA coating and etching conditions, but with the substrate held flat during etching.

We then image both PMMA-covered sensor configurations using a 20× objective and record the transmitted intensity with a near-infrared camera while scanning the illumination wavelength across the spectral range spanning both the differential ($\nabla n$) and refractometric ($\bar{n}$) modes. An unpatterned blank region adjacent to the metasurface area provides a direct reference. As shown in Fig. 5c, the sensor area exhibits reduced transmittance at sensing wavelengths due to resonant reflection, producing an intensity contrast with the blank region. The experimental imaging results are shown in Fig. 5d,e. For the tilted PMMA coverage, the recorded images reveal clear contrast between the metasurface and adjacent blank regions at both sensing wavelengths (Fig. 5d). By contrast, laterally uniform PMMA coverage produces a resonant signal only at the refractometric mode, with no discernible inhomogeneity signature (Fig. 5e).

The corresponding transmittance spectra for the two PMMA configurations are shown in Fig. 5f. Consistent with the imaging results, a distinct resonance appears near 1.477 μm only for the slanted PMMA perturbation, whereas the refractometric resonances for both configurations nearly overlap. We attribute the reduced contrast of the inhomogeneity spectral signature to aperiodic variations along the tilted PMMA interface, arising from local PMMA thickness and etch-rate variations that disrupt coherent lattice coupling. As evident in the AFM image in Fig. 5b, the slanted profiles exhibit nonuniform curvature and surface roughness that vary across multiple unit cells. The resulting resonant damping effect is more pronounced for the high-$Q$ resonances associated with small transverse variations. We further tune the transverse variation by changing the substrate tilt angle during oblique etching. As shown in Fig. 5g, the spectra around the inhomogeneity resonance exhibit progressively reduced extinction as the tilt angle decreases, consistent with stronger radiative damping of the quasi-BIC.

In summary, we demonstrate an inverse-designed metasurface sensor capable of resolving refractometric loading and spatial inhomogeneity through two spectrally distinct channels. This new sensing capability could enable broader studies of complex biological and material systems, with potential applications including probing nanoscale compositional heterogeneity and the evolution of spatial organization[54–57]. Although the nonlocal quasi-BIC presently requires a periodically repeated transverse perturbation to preserve coherent excitation, future designs that localize this mode could accommodate more general analyte distributions, albeit with reduced $Q$ factor and sensitivity. Multilayer architectures[58–62] or independently designed sensing arrays[15,27] could further expand the accessible sensing functionalities across different spatial scales and spectral ranges.

More broadly, the emergence of an untargeted modal response in our device suggests an exploratory role for inverse design beyond prescribed optimization objectives. By exploiting optimization as a discovery tool, inverse design could uncover unanticipated optical functionalities that retain certain characteristic modal features encoded by complex freeform nanostructures, potentially opening new opportunities for controlling light–matter interactions in spontaneous-emission control[63–66], nonlinear optics[67–70], optomechanics[71,72], and photochemistry[73,74].

## Methods

### Fabrication

The metasurfaces were fabricated from a 462-nm-thick amorphous-silicon film deposited on fused silica by plasma-enhanced chemical vapor deposition. A bi-layer electron-beam resist of PMMA 495K A4 and PMMA 950K A4 was sequentially spin-coated at 4,500 r.p.m., followed by electron-beam evaporation of a 10-nm chromium layer as a conduction layer. The freeform patterns were then defined by electron-beam lithography. After development, a 15-nm-thick $Al_2O_3$ layer was deposited by electron-beam evaporation and used as a hard mask for subsequent pattern transfer. The silicon nanostructures were etched by reactive ion etching using a mixed gas of $SF_6$ and $C_4F_8$. The PMMA etching processes shown in Fig. 5 were performed using pure oxygen at a flow rate of 20 sccm, with RF and ICP powers of 20 and 500 W, respectively, under a chamber pressure of 30 mTorr.

### Measurement

The device was illuminated with an *x*-polarized plane wave from a supercontinuum source (NKT SuperK EVO). The transmitted light was collected using a microscopy system comprising a 20× objective (Mitutoyo) and a tube lens, and subsequently divided by a beam splitter into two detection paths. One arm was imaged with a near-infrared camera (Wildcat+ 640), while the other was directed to a high-resolution spectrometer (HORIBA iHR320) for spectral characterization.

**Full-wave numerical analysis**

Full-wave numerical simulations of the all-dielectric metasurface were performed and cross-validated using the open-source finite-difference time-domain package MEEP, the commercial finite-difference time-domain solver Lumerical FDTD, and the finite-element-method solver COMSOL Multiphysics. The refractive indices of $SiO_2$ and PMMA were set to 1.45 and 1.48, respectively, and the silicon index (3.54 near 1600 nm) was obtained using ellipsometry. The spectra and field distributions presented in this work were calculated using a single meta-unit with periodic boundary conditions in the lateral directions. An *x*-polarized plane wave was incident normally from the substrate side. The maximum mesh dimension was maintained below one-tenth of the shortest wavelength considered in the simulations.

**Data Availability**

The data that support the plots within this paper and other findings of this study are available from the corresponding authors upon reasonable request.

**Acknowledgments**

Y.Z. acknowledges support from UNC Charlotte start-up funds and National Science Foundation (NSF) under award ECCS-2501853. The silicon thin film growth was conducted at the Center for Nanophase Materials Sciences, which is a DOE Office of Science User Facility. The remainder of the fabrication process took place in the cleanroom facilities of the Center for Optoelectronics and Optical Communications at UNC Charlotte. We thank Adam Overvig for fruitful discussions.

**Contributions**

Y.Z. conceived the idea. Z.J. conducted the design, modeling, theoretical analysis and experimental measurements. Z.J. and N.G. prepared the gradient-profile analyte samples and performed the associated numerical analysis and measurements. S.S. fabricated the samples. Y.Z. and Z.J. wrote the manuscript with input from all authors. Y.Z. supervised the project.

**Competing interest**

The authors declare no competing interests.

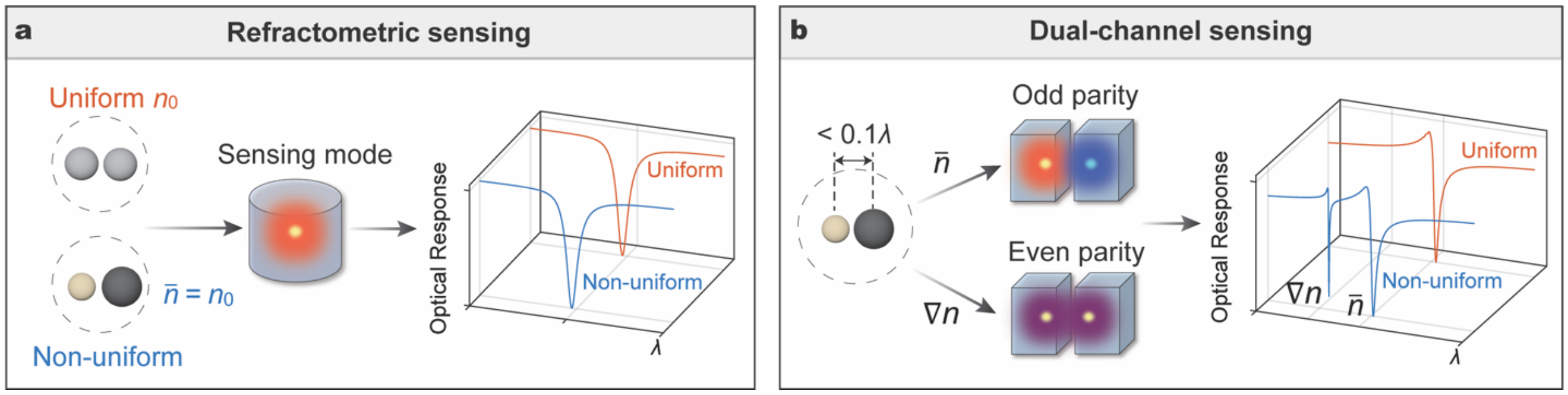

**Figure 1 | Dual-channel nanophotonic sensing.** **a,** Conventional refractometric sensing. Uniform and inhomogeneous analyte distributions with similar mean indices produce nearly identical responses in a single spectroscopic channel. **b,** Dual-channel refractometric and inhomogeneity sensing. The sensor supports two high-quality-factor ($Q$) modes with deeply structured near-field profiles that respond to complementary analyte perturbations, encoding mean refractometric loading ($\bar{n}$) and transverse inhomogeneity ($\nabla n$) into two spectrally resolvable channels.

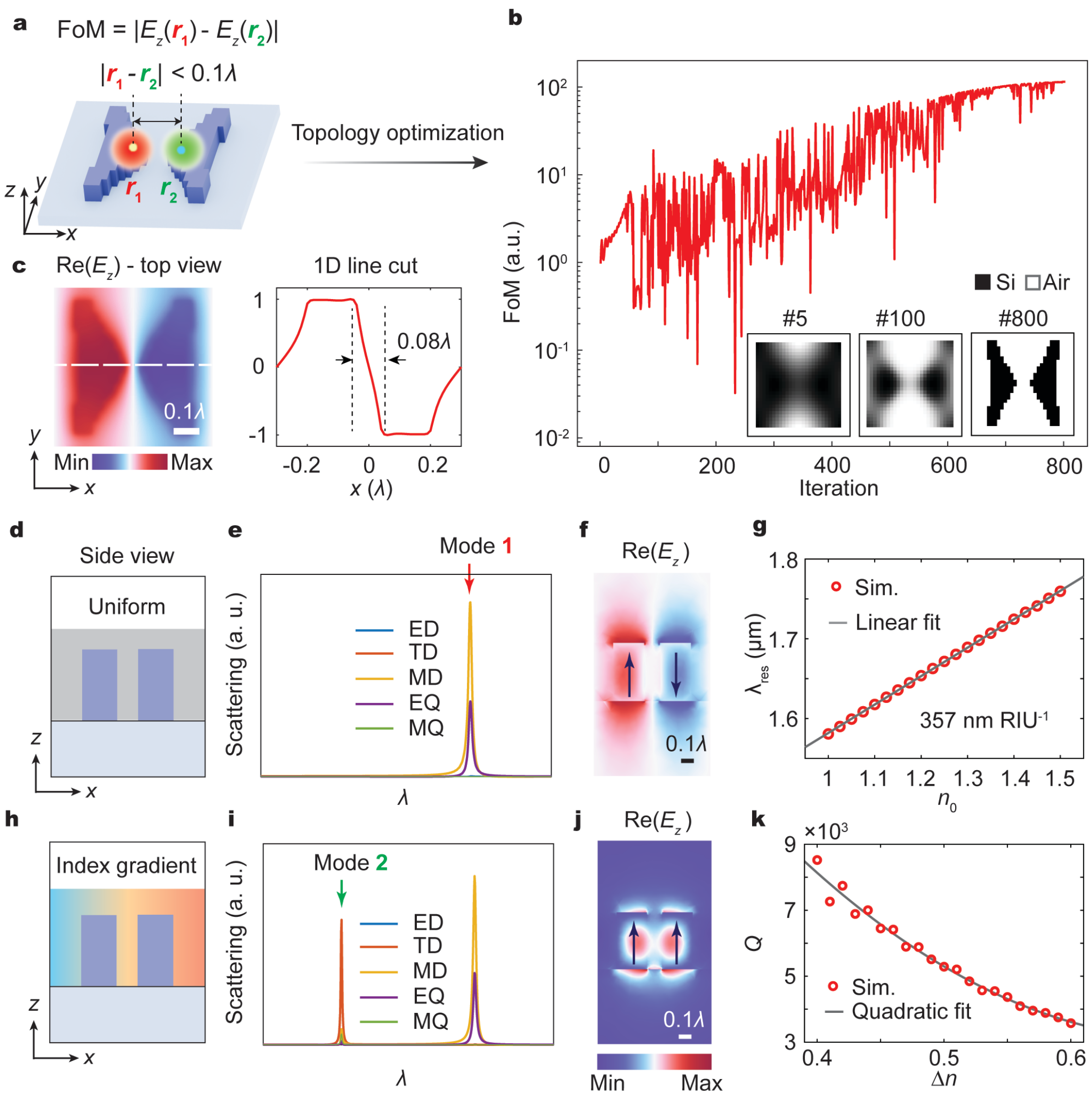

**Figure 2 | Design and analysis of the dual-channel metasurface sensor. a,** Inverse-design framework. The figure of merit (FoM), defined as $|E_z(\boldsymbol{r_1}) - E_z(\boldsymbol{r_2})|$, targets lateral field compression by maximizing the field contrast between two closely spaced probes (<λ/10). **b,** Optimization trajectory. Insets: structural evolution within the unit cell. **c,** Top-view $\mathrm{Re}(E_z)$ distribution (left) and corresponding one-dimensional line cut (right), revealing a field reversal confined to a 0.08λ-wide region. **d,** Refractometric sensing configuration with a uniform dielectric overlayer. **e,** Multipolar decomposition of the analyte-loaded metasurface. ED: electric dipole, TD: toroidal dipole, MD: magnetic dipole, EQ: electric quadrupole, MQ: magnetic quadrupole. **f,** Side-view $\mathrm{Re}(E_z)$ distribution at resonance, showing the odd-parity antiparallel field configuration with respect to the central mirror plane. **g,** Refractometric response quantified by tracking the resonance wavelength with increasing superstrate index. **h,** Inhomogeneity-sensing configuration with a linearly varying transverse index profile. **i,j,** Multipolar decomposition (**i**) revealing an additional resonance with an even-parity field distribution (**j**). **k,** Inhomogeneity response quantified by the $Q$ factor as a function of transverse index contrast $\Delta n$.

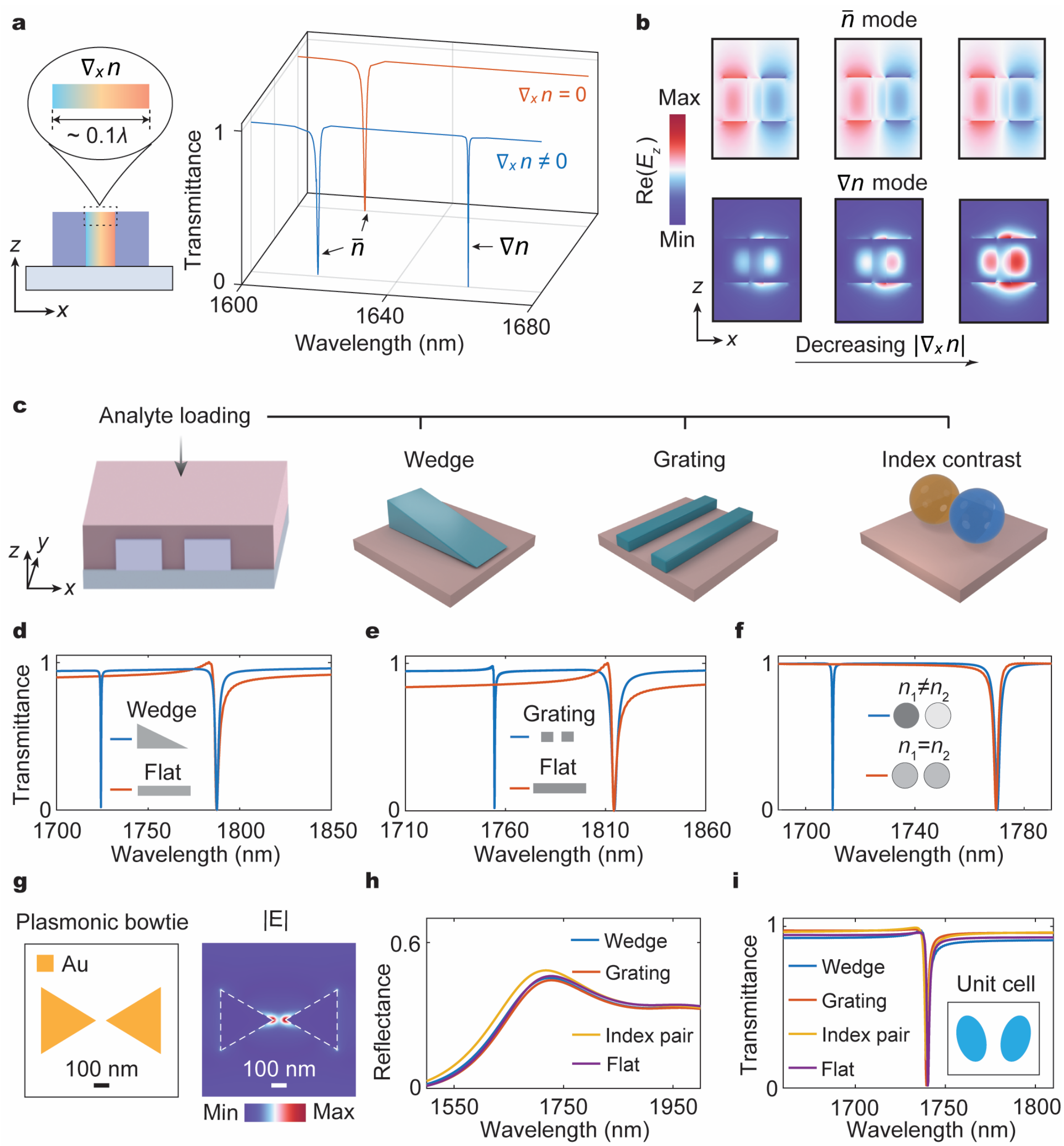

**Figure 3 | Deep-subwavelength spatial inhomogeneity sensing. a,** Analyte confined to the ~λ/10-wide central gap with a linearly varying transverse index profile (left). Transmittance spectra for laterally uniform ($\nabla_x n = 0$) and varying ($\nabla_x n \neq 0$) analyte profiles with the same mean index (right). **b,** Side-view $Re(E_z)$ distributions at the two sensing resonances for various analytes with fixed mean index and varying transverse gradient. **c,** Inhomogeneity sensing for diverse nanoscale analyte configurations. A planar polymer coating provides the analyte-loading surface. Three representative cases are considered: (1) a dielectric wedge with an inclined interface, (2) a nanoscale grating and (3) a pair of nanospheres with different refractive indices. **d-f,** Transmittance spectra for the three inhomogeneous cases and their respective controls. **g-i,** Comparative sensing analysis using representative refractometric platforms: a plasmonic bowtie antenna with a localized gap hotspot (**g,h**) and a dielectric zigzag array of anisotropic resonators (**i**).

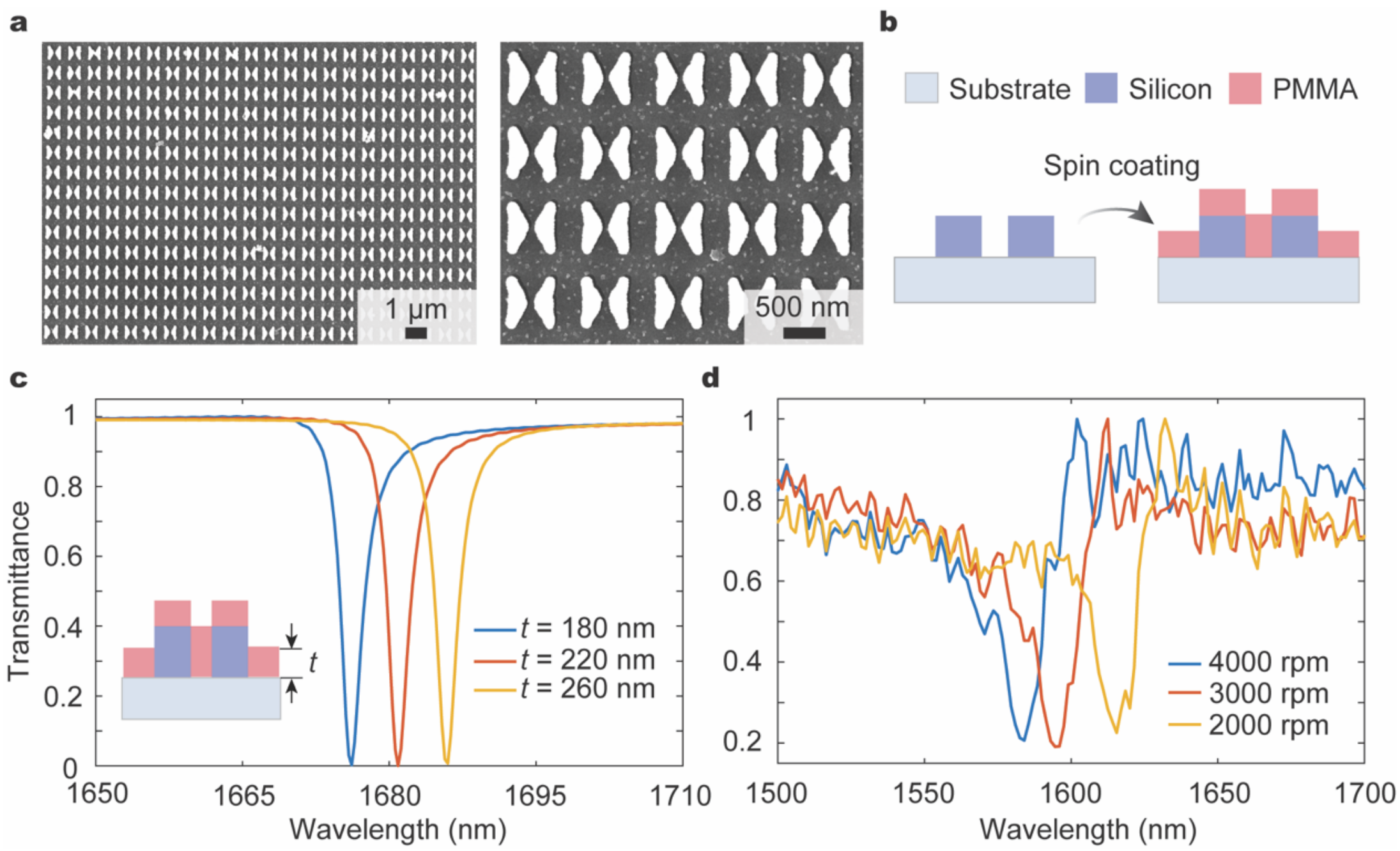

**Figure 4 | Experimental demonstration of the refractometric sensing channel. a,** Scanning electron microscopy images of the fabricated metasurface sensor. **b,** Tunable analyte loading through spin coating of PMMA layers with different thicknesses. **c,** Simulated transmittance spectra for PMMA thicknesses ranging from 180 to 260 nm. **d,** Experimental transmittance spectra for different PMMA thicknesses obtained by varying the spin-coating speed from 2,000 to 4,000 r.p.m.

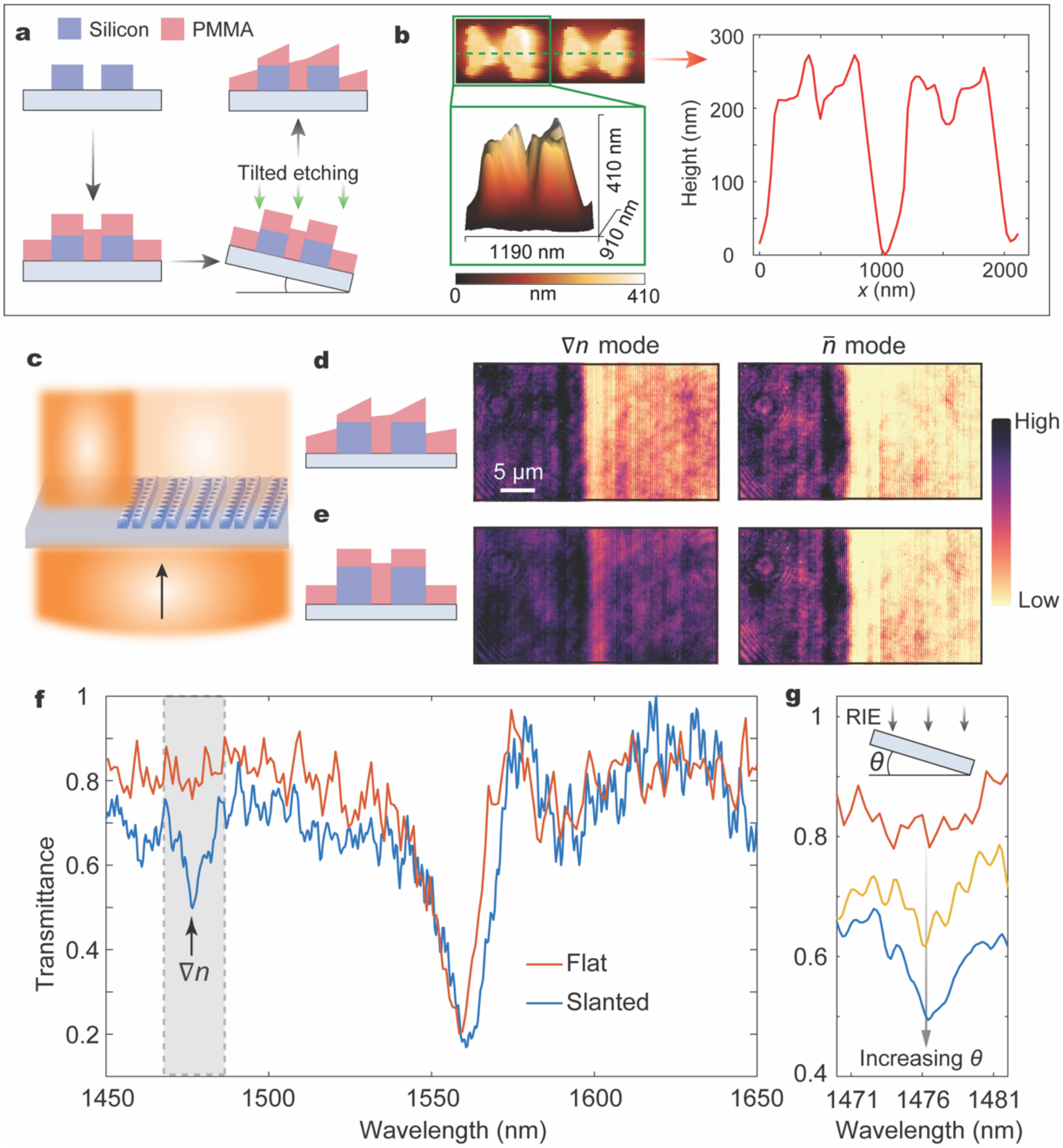

**Figure 5 | Experimental demonstration of the inhomogeneity-sensing channel. a,** Fabrication flow for creating transversely varying analyte coverage. A planar PMMA film is first deposited over the nanostructures, followed by tilted reactive ion etching to form a slanted PMMA interface. **b,** Atomic force microscopy image of the PMMA-covered sample after tilted etching, with the corresponding three-dimensional rendering and central line profile (green dashed line). **c,** Schematic of imaging scheme at differential ($\nabla n$) and refractometric ($\bar{n}$) resonances. An adjacent unpatterned region serves as a reference for visualizing resonant contrast. **d,e,** Experimental transmittance images for tilted (**d**) and laterally uniform (**e**) PMMA coverage. **f,** Measured transmittance spectra for the two PMMA configurations. **g,** Zoomed-in spectra around the inhomogeneity resonance for samples with different transverse variations, generated by varying the substrate tilt angle during oblique etching (inset).